\documentclass[reprint,12pt]{JASA}

\usepackage{algpseudocode}

\begin{document}

\title[A Time-Scale Modification Dataset]{A Time-Scale Modification Dataset with Subjective Quality Labels}
\author{Timothy Roberts}
\author{Kuldip K. Paliwal}
\affiliation{Signal Processing Laboratory, Griffith University, 170 Kessels Road, Nathan, QLD 4111, Australia}
\email{timothy.roberts@griffithuni.edu.au}

\preprint{Roberts, Griffith University}		%  if you want want this message to appear in upper left corner of title page

\date{\today}

\begin{abstract}
Time-Scale Modification (TSM) is a well-researched field, however no effective objective measure of quality exists.  This paper details the creation, subjective evaluation and analysis of a dataset, for use in the development of an objective measure of quality for TSM.  Comprising two parts, the training component contains 88 source files processed using six TSM methods at 10 time-scales, while the testing component contains 20 source files processed using three additional methods at four time-scales.  The source material contains speech, solo harmonic and percussive instruments, sound effects and a range of music genres. 42,529 ratings were collected from 633 sessions using laboratory and remote collection methods.  Analysis of results shows no correlation between age and quality of rating; expert and non-expert listeners to be equivalent; minor differences between participants with and without hearing issues; and minimal differences between testing modalities.  Comparison of published objective measures and subjective scores shows the objective measures to be poor indicators of subjective quality.  Initial results for a retrained objective measure of quality are presented with results approaching average root mean squared error loss and Pearson correlation values of subjective sessions. The labelled dataset is available at http://ieee-dataport.org/1987.
\end{abstract}

%% pacs numbers not used

\maketitle

%  End of title page for Preprint option --------------------------------- %

\section{\label{sec:Introduction} Introduction}
% \citep{Flanagan_Golden_1966, Portnoff_1976, Roucos_Wilgus_1985, Verhelst_Roelands_1993, Laroche_Dolson_1999, Driedger_Muller_Ewert_2014, Damskagg_2017, Sharma_2017, Roberts_2019_FESOLA}
% No current objective measures
% Need Subjective results to develop an objective measure
% details first objective measure
% Note that more objective measures will be reported.

Time-Scale Modification (TSM) is the process of modifying the duration of a signal without modifying timbre and pitch.  It has found use in areas including music production, language learning and speech recognition systems.  Despite being a well-researched field, an effective objective measure of quality has not yet been published, limiting comparisons between TSM algorithms. When subjective evaluation has been used, each paper has used a unique set of source material and methods, further reducing comparison to only the methods involved in the evaluation.  In order to develop an effective objective measure, a dataset with subjective quality labels is required.  This work details the creation, subjective evaluation and analysis of the first dataset for this purpose, and gives preliminary results for a neural-network-based objective measure of quality.

TSM algorithms most commonly modify the temporal domain by varying the ratio between analysis ($S_a$) and synthesis ($S_s$) shift sizes within an Analysis Modification Synthesis framework. This ratio, given by
\begin{equation}
  \label{eq:alpha}
    \beta = \frac{1}{\alpha}=\frac{S_a}{S_s}
\end{equation}
shows $\alpha$ to be the change in signal duration \citep{Roucos_Wilgus_1985}, while $\beta$ is the playback speed \citep{Sylvestre_Kabal_1992} and will be used within this paper.

Algorithms for TSM can be classified into three main categories: frequency domain, time domain and hybrid methods. In general, frequency-domain methods excel in scaling harmonically complex material but struggle to produce high quality results with highly transient signals.  Time-domain methods are more effective at scaling transient signals but give poor results for polyphonic signals.  Hybrid methods leverage the strengths of frequency and time domain methods to produce higher quality results \citep{Driedger_Muller_Ewert_2014}.

Common artefacts produced during TSM include `phasiness' and reverberation \citep{Portnoff_1981, Laroche_Dolson_1997}, musical and metallic noise or undesirable roughness \citep{Laroche_Dolson_1999}, a buzzy quality \citep{Laroche_2002} and transient smearing \citep{Laroche_Dolson_1999}.  Phasiness and reverberation are heard as a loss of spectral definition and are most commonly associated with frequency domain methods.  \citet{Laroche_Dolson_1999} suggest that this is due to a change in relationship between the phases of bins in the spectral domain.  Musical noise, also known as musical artefacts or musical tones, is due to isolated holes and/or peaks within the power spectrum \citep{Matteo_2019}.  Within TSM, these artefacts are caused by periodicity introduced to noise bins during phase progression, due to the sum of sines model of the Short Time Fourier Transform (STFT).  Depending on the frequency relationships between these periodic signals the noise will be perceived as musical for simple harmonic relationships and metallic for complex harmonic relationships.  Transient smearing occurs due to the trade off between STFT spectral and temporal resolution in frequency domain algorithms.  As the frame size increases to improve spectral resolution, temporal resolution decreases leading to smearing of transients in time. The buzzy quality, also known as transient skipping or duplication, is an artefact of time-domain methods in which transients may be skipped for $\beta>1$ or duplicated for $\beta<1$.

The aim of TSM is often noted, however an exploration of ideal TSM has not been published. For the purpose of subjective evaluation, we describe ideal TSM as indistinguishable from a change by the sound source, that is: the processing should be transparent.  A musician changing tempo or a speaker changing cadence would therefore be ideal and should be the goal for TSM algorithms.  Consequently, ideal TSM should be determined by the sound source being scaled.  For example, a dry recording of individual clicks simply requires temporal realignment of each click, however a recording of sustained notes played on a violin would require the extension of the sustain section of the note's envelope.  Further, in the case of a piano, one must consider whether the transient or harmonic nature of the source should be maintained.  If a staccato melody played in the upper register without damping is to be slowed, should note decay be lengthened or should the decay be maintained with each note shifted to the new time-scale?  We argue that as the piano is a percussive instrument and unable to modify its amplitude envelope, the note decay should be maintained.  This is counter to the processing applied by almost all published TSM algorithms.  We propose that an ideal TSM algorithm would be sensitive to the signal source and be capable of modifying only the sustain portion of the amplitude envelope.  This raises many questions in the processing of reverberation, vibrato, specific phonemes and more.  We consider that content aware or source sensitive TSM is an area with considerable potential for improving the quality of TSM.

% Reviewer 2 - Consider for example a melody played in the upper register of a piano, where tones decay quickly. When TSM is applied to slow it down, the decay time of the piano tones is also increased. Clearly, it is not the same as playing that melody in a slower tempo with that piano.

The remainder of the paper is laid out as follows.  Section~\ref{sec:Algorithms} describes the TSM algorithms used to create the dataset and previous methodologies for quality evaluation.  Section~\ref{sec:Dataset} describes the source files used in the creation of the dataset and the processing of the source material to create the processed dataset.  Section~\ref{sec:Subjective_Testing} describes the subjective testing methodology, opinion score normalization, results and analysis of the subjective testing and dataset availability.  Section~\ref{sec:Objective} compares subjective results with published objective measures and provides preliminary results for an novel objective measure of quality.  Finally, section~\ref{sec:Conclusion} summarises and draws conclusions from this research.

% ----------------------------------Time-Scale Modification Algorithms----------------------------------
\section{Algorithms and Quality Evaluation}
\label{sec:Algorithms}

% \subsection{Phase Vocoder}
The Phase Vocoder (PV), is a frequency-domain method that uses the known phase progression between frames at the original time-scale to calculate the phase progression between frames at the adjusted time-scale.  The digital implementation by \citet{Portnoff_1976} uses the STFT to calculate phase spectra and forms the basis for all PV methods published since.  The PV is effective at scaling signals with a complex harmonic structure, however it introduces `phasiness' for non-integer values of $\alpha$ and is prone to transient smearing. See \citet{Laroche_Dolson_1999} for detailed explanation.

% \subsection{Identity Phase Locking Phase Vocoder}
The Identity Phase Locking Phase Vocoder (IPL) \citep{Laroche_Dolson_1999} reduces 'phasiness' introduced by the PV algorithm.  The PV maintains horizontal phase coherence within each STFT bin, however the vertical phase coherence between bins is not maintained.  In IPL, the phase of magnitude spectrum peaks are modified, with nearby bins locked to the phase progression of the closest peak.  This method was extended, through multi-resolution peak-picking and accounting for added or removed peaks by \citet{Karrer_Lee_Borchers_2006}.  These methods reduce phasiness, however they can introduce a spectral roughness known as metallic or musical noise.

% \subsection{Waveform Similarity Overlap-Add}
The Waveform Similarity Overlap Add algorithm (WSOLA) \citep{Verhelst_Roelands_1993} is a time-domain method that uses the similarity between a frame and its natural progression in the input signal to minimize discontinuities in the time-scaled signal.  This is in contrast to previous methods that compare with the output signal \citep{Roucos_Wilgus_1985,Moulines_1990}.  WSOLA effectively processes speech and monophonic musical signals, however due to the reliance on the fundamental frequency for alignment, produces low quality results for polyphonic signals.

% \subsection{Fuzzy Epoch Synchronous Overlap-Add}
Fuzzy Epoch Synchronous Overlap-Add (FESOLA) \citep{Roberts_2019_FESOLA} uses cross-correlation of glottal closure instants, known as epochs, for aligning frames of speech. Epochs are calculated using a Zero Frequency Resonator before smearing in the time-domain.  The smearing improves the cross-correlation of epochs, and accounts for changes in fundamental frequency.  This method works well for speech and monophonic signals, however it is not effective at processing polyphonic signals.

% \subsection{Harmonic-Percussive Separation Time-Scale Modification }
Harmonic-Percussive Separation Time-Scale Modification (HPTSM) of \citet{Driedger_Muller_Ewert_2014} is a hybrid method that uses median filtering of spectrograms for signal separation.  WSOLA and IPL are used for percussive and harmonic components respectively.  Improved quality was shown over both individual methods.  The method was also shown to compete with contemporary commercial state-of-the-art algorithms.

% \subsection{Multi-component Time-Varying Sinusoidal Decomposition}
Multi-component Time-Varying Sinusoidal decomposition (uTVS) \citep{Sharma_2017} uses a Mel-scale filter-bank and the Hilbert transform to calculate instantaneous phase and frequency, bypassing phase unwrapping and the quasi-stationary assumption of traditional frequency-domain methods.  As a result, temporal smearing and `phasiness' artefacts are reduced.  This method slightly improves quality over HPTSM, with large improvements over traditional methods.

% \subsection{Zplane Elastique}
Elastique \citep{elastique} is a widely used commercial TSM method.  While the algorithm is not publicly available, it is currently a state-of-the-art method and has been used in recent TSM subjective evaluations.

% \subsection{Phase Vocoder with Fuzzy Classification of Spectral Bins}
Fuzzy classification of spectral bins (FuzzyPV) \citep{Damskagg_2017}, is an extension of the IPL.  Spectral bins are given a degree of membership to three classes, sinusoidal, noise and transient, resulting in a fuzzy classification of each bin.  Sinusoidal bins are scaled using IPL with phase locking applied to sinusoidal bins, while random phase is added to noise bins.  Analysis phases of transients bins are simply relocated in time.  Subjective evaluation shows improvement over HPTSM and similar performance to Elastique.

% \subsection{Non-Negative Matrix Factorization}
Non-Negative Matrix Factorization Time-Scale Modification (NMFTSM) by \citet{Roma_Green_Tremblay_2019} decomposes the signal into percussive events and harmonic components.  Percussive events are copied directly to the output signal, while IPL is used for harmonic components.  The duration of percussive events is preserved, however it is highly reliant on correct detection of the events and introduces novel artefacts.

% ----------------------------------{Prior Quality Assessment Methodologies----------------------------------
% \section{Prior Quality Assessment Methodologies}
% \label{sec:Prior_Quality_Methodologies}
% \citep{Driedger_Muller_Ewert_2014, Damskagg_2017, Sharma_2017, Roberts_2019_FESOLA, Karrer_Lee_Borchers_2006, Moinet_Dutoit_2011, Jun_2007, Roberts_2018_Stereo}
% \citep{Roucos_Wilgus_1985, Verhelst_Roelands_1993, Laroche_Dolson_1999, Puckette_1995, Bonada_2000, Dorran_2004, Kafentzis_2013, Roberts_2018_FDTSM, Roma_Green_Tremblay_2019}
Little formal subjective testing has been used to evaluate proposed methods, with most proposed methods providing results from informal testing.  A wide variety of time-scales and algorithms are used, with little consistency.  Time-scales are often limited with two to five times scales ($0.5\leq \beta \leq 2$) reported in formal testing, with a bias towards $\beta<1$.  This reduces the number of files that require rating, but also limits algorithm evaluation.  The difference in quality between $\beta<1$ and $\beta>1$ was mentioned briefly by \citet{Sylvestre_Kabal_1992}.  Since the release of the MATLAB TSM Toolbox \citep{Driedger_Muller_2014}, PV, IPL, WSOLA and HPTSM, have been used in most evaluations, while comparisons to commercial algorithms are rare \citep{Karrer_Lee_Borchers_2006, Driedger_Muller_Ewert_2014, Damskagg_2017}.  The source audio used during testing also varies between papers with some papers using the files provided with the MATLAB TSM Toolbox.  It was noted by \citet{Moulines_Laroche_1995} that a thorough perceptual evaluation of TSM approaches had not yet been undertaken.

Two objective measures have been proposed, Signal to Error Ratio ($SER$) by \citet{Roucos_Wilgus_1985} and synthesis consistency ($D_M$) by \citep{Laroche_Dolson_1999}.  $SER$ accounts only for successive magnitude spectra, with no attention paid to phase spectra.  $D_M$ also compares the output frame's magnitude to the reconstructed signal's magnitude, however the ``measure is not a clear indicator of phasiness'' \citep{Laroche_Dolson_1999}.  Neither of these measures have seen continued use.

% ----------------------------------Dataset Source Material----------------------------------
\section{Dataset Description}
\label{sec:Dataset}
% \subsection{Source Material Dataset}
% \label{sec:Dataset_Source}
The source material for the dataset was collated from the author's previous creative projects including films, concert and field recordings as well as music written specifically for the dataset.  Files were selected to give a broad spectrum of content with variation in TSM difficulty.  The number of source files, methods and time-scales was determined by balancing the amount of content required to train a neural network and the number of ratings required for a `true' Mean Opinion Score (MOS). All content was converted to mono by averaging each pair of samples to remove the influence of poor handling of multi-channel files \citep{Roberts_2018_Stereo} and normalized to $\pm$1 before TSM.  All files are 16-bit with a sample rate of 44.1kHz and range in SPL from 56.62dB to 86.92dB with a mean and standard deviation of 73.37dB and 6.75dB.

The full dataset contains 34 musical, 37 solo instrument and 37 voice files with a complete listing provided with the dataset.  The total playback length of the source files is 6 minutes and 42 seconds. Duration was kept short, with a mean of 3.7 seconds and standard deviation of 1.6 seconds, to limit the duration after time-scaling.  Files were recorded using a combination of close microphone placement, multi-microphone concert recording, digital synthesis and sampling techniques and shotgun, lapel and large diaphragm condenser microphones.  These variations in source material allow for extended subjective evaluation of future TSM methods.  The musical and solo files contain synthetic and organic sound sources across classical, rock, jazz, and electronic genres.  Voice files contain singing and male, female, and child speech.  Finally, the evaluation source files contain a mix of each file type and were used in the generation of the test and evaluation subsets.  Table \ref{tab:Files} shows an overview of the signal sources.

\begin{table*}[ht]
\caption{Signal sources in each dataset class. Sources considered are Total, Brass, Percussion, Piano, Rhythm Section, Sound Effects, Strings, Synthesizers, Woodwinds, Child, Female, Male and Singing.  All sources within a file are counted separately.}
\centering
\begin{ruledtabular}
\begin{tabular}{cccccccccccccc}
 & \textbf{Total} & \textbf{Br.} & \textbf{Perc.} & \textbf{Piano} & \textbf{Rhythm} & \textbf{SFX} & \textbf{String} & \textbf{Synth.} & \textbf{Wood.} & \textbf{Ch.} & \textbf{F.} & \textbf{M.} & \textbf{Sing.} \\
\hline
\textbf{Music} & 27 & 6 & 7 & 6 & 8 & 2 & 3 & 9 & 12 & - & - & 1 & 2 \\
\hline
\textbf{Solo} & 31 & - & 11 & 3 & 4 & 1 & 1 & 3 & 11 & - & - & - & - \\
\hline
\textbf{Voice} & 30 & - & - & - & - & - & - & - & - & 3 & 12 & 15 & 4\\
\hline
\textbf{Eval} & 20 & 1 & 2 & 2 & 3 & 1 & 1 & 2 & 9 & 1 & 3 & 3 & - \\
\end{tabular}
\end{ruledtabular}
\label{tab:Files}
\end{table*}

% The musical files consist of six synthetic music excerpts, 17 organic excerpts and three excerpts containing a mix of synthetic and organic sound sources. This classification contains examples of classical, rock, jazz, and a variety of electronic genres.  Six files contain brass instruments, seven contain percussion, four contain piano, four contain a rhythm section, three contain stringed instruments, five contain synthesizers and 10 contain woodwind instruments.  The solo instrument files consist of five synthetic and 25 organic instruments. 11 files contain percussion, four contain rhythm instruments, one contains violin, three contain synthesizers and 11 contain woodwind instruments.  The voice files can be further classified, with 14 male, eight female, three child and five singing files.  Finally, the objective source files contain a mix of the aforementioned file types and are used in the generation of the test and evaluation sets.

% ----------------------------------Database Processed Material----------------------------------
% \subsection{Processed Material Dataset}
% \label{sec:Dataset_Processed}
To form the training set, the source dataset was processed using the first six methods previously mentioned at 10 time-scale ratios resulting in 5,280 processed files.  Time-scale ratios of 0.3838, 0.4427, 0.5383, 0.6524, 0.7821, 0.8258, 0.9961, 1.381, 1.667, and 1.924 were generated randomly, but adjusted to ensure coverage across the range of interest.  The testing set used Elastique, FuzzyPV and NMFTSM at four random time scales in four bands across $0.25\leq \beta \leq 2$, resulting in 240 testing files.  Subjective evaluation was conducted for both the training and testing sets.  An additional evaluation set was created and is discussed in section \ref{sec:Objective}. Full dataset generation took approximately three days on a medium to high end workstation.

%\citep{Murty_2008} ZFR reference
The MATLAB TSM Toolbox \citep{Driedger_Muller_2014} was used with default settings for WSOLA, HPTSM and Elastique time-scaling.  FuzzyPV and NMFTSM used provided implementations with default settings.  Author implementations of PV, IPL, uTVS and FESOLA were used with Hann windowing throughout and parameters chosen to maximize informal subjective evaluation. All files were normalized after processing.  The PV and IPL used a frame length of 2,048 samples (46.4ms) and synthesis hop of 512 samples.  FESOLA used a frame length of 1024 samples (23.2ms).  WSOLA used a frame length of 1,024 samples (23.2ms) a synthesis hop of 512 samples and a tolerance of 512 samples.  HPTSM used identical IPL parameters while WSOLA had a frame size of 256 samples (5.8ms) and a synthesis hop of 64 samples.  uTVS was implemented using six times oversampling and a filterbank containing 88 filters to maintain the relationship between the signal sample rate and filterbank length of the original paper.  During testing, an error in the uTVS implementation was found that introduced discontinuities within spectra during processing at $0.9\leq \beta \leq 1.1$ for some files.  However, as the purpose of the subjective testing was to rate multiple files with a variety of artefacts, they were not removed from the dataset.  The error was rectified before creation of the evaluation subset.

% ----------------------------------Subjective Testing Methodology----------------------------------
\section{Subjective Testing}
\label{sec:Subjective_Testing}
\begin{figure}[ht]
    \centering
    \includegraphics[width=\reprintcolumnwidth]{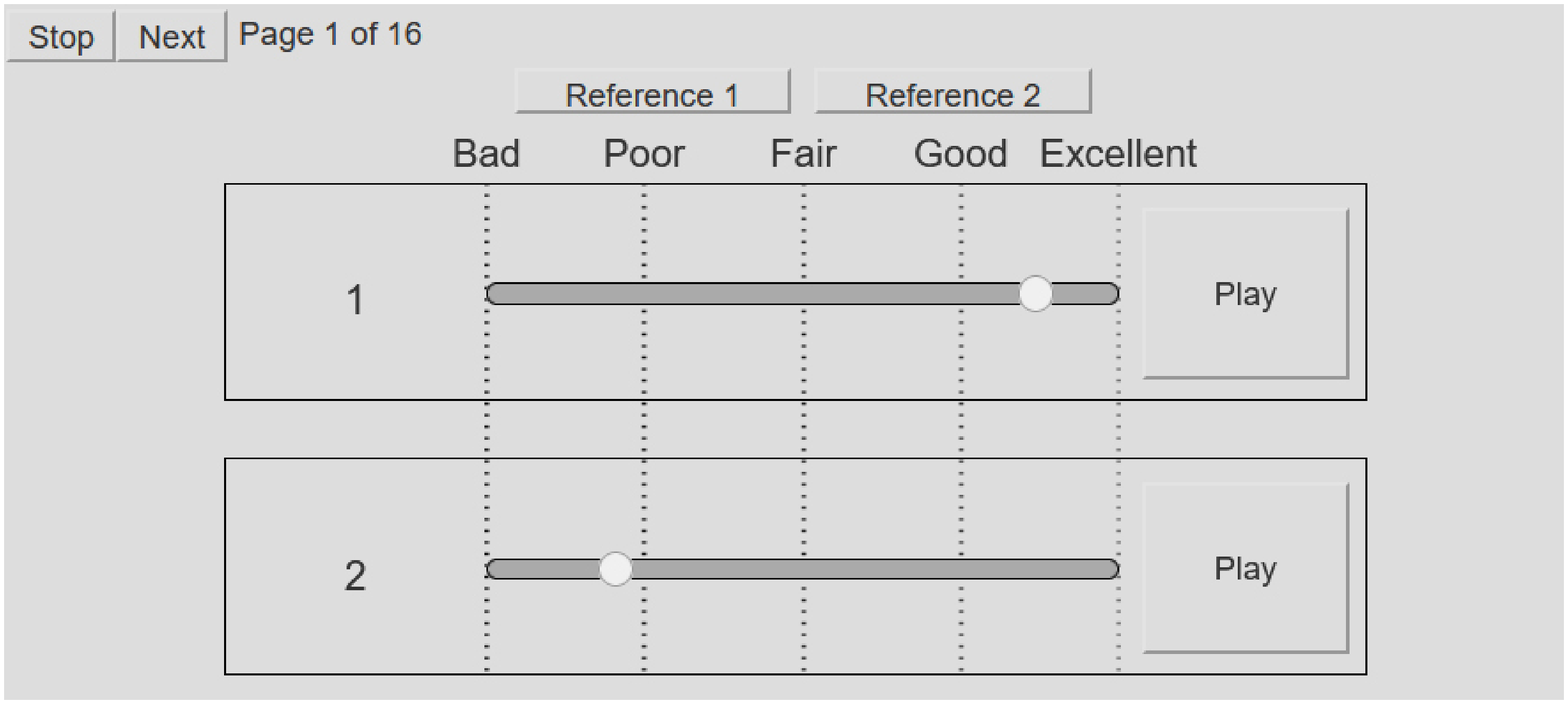}
    \caption{Web Audio Evaluation Tool user interface used for remote testing. Shown with two file pairs.}
    \label{fig:Remote_GUI}
\end{figure}

Subjective testing was undertaken in two phases.  Initial testing was conducted internally within the laboratory.  Due to the large number of responses needed per file, testing transitioned to an online browser-based test using the Web Audio Evaluation Tool (WAET) \citep{WAET_2015}, shown in figure \ref{fig:Remote_GUI}. Remote testing greatly increased the number of participants in the study.  Participants were contacted in person, directly through social media and email, through mailing lists and public posts on websites such as Reddit and Facebook.

Testing followed ITU-R BS.1248-1 \citep{ITU_BS1284} recommendations for general methods for the subjective assessment of sound quality as close as practicable, resulting in the following testing parameters.  Files were presented in reference-processed pairs with no limits placed on the amount of playback before moving to the next file.  Checks were included to ensure both files were played at least once.  A continuous grading scale was used in conjunction with a quality scale, where Poor-Excellent corresponds to scores of 1-5.  Sessions contained a randomised selection of processed files, presented in random order, with participants free to choose the session they would evaluate.  The amount of content per session was refined during testing, for a maximum session duration of 20 minutes.  Towards the end of testing, the sessions were restricted to files that had limited responses to reduce MOS standard deviation.

% \subsection{Laboratory testing}
Initial testing was undertaken using a bespoke MATLAB GUI that presented individual reference-processed pairs, allowed for saving and restoring of sessions, user input of name, sound transducer, and a check that the participant had no known hearing issues.  Participants received training before beginning testing, including explanations of the purpose of TSM and common artefacts with audio examples.  A small initial test session of 33 files was completed before a random session was assigned.  Each session contained 18 minutes of audio, approximately 200 files, randomly selected from the pool of processed audio files.  Participants could elect to evaluate additional sessions following a break equal in length to the completed session.

% \subsection{Remote testing}
To increase the number of participants, the WAET was used.  A small number of sessions were evaluated containing 100 files before reduction to 60 files based on participant feedback of session duration.  Training identical to laboratory testing was available from the index page, which contained links to each test session.  The index page contained reminders to use headphones in a quiet space during testing and a random number generator to suggest which test session the participant should complete.  Before each session, name, age, sound transducer, experience in critical evaluation of sound and any known hearing issues were collected.  Participants could also elect provide an email address to be contacted for future studies.  Each session was split into pages containing six reference-processed pairs.

% \subsection{Session Normalization}
To remove bias and variability between sessions, opinion scores were normalized according to ITU-R BS1284 \citep{ITU_BS1284} using
\begin{equation}
 \label{eq:MOS_Norm}
 Z_i = \frac{x_i-\bar{x}_{si}}{\sigma_{si}}\sigma_s+\bar{x}_s
\end{equation}
where $Z_i$ is the normalized result, $x_i$ is the opinion score of subject $i$, $\bar{x}_{si}$ is the mean score for subject $i$ in session $s$, $\bar{x}_s$ is the mean score of all subjects in session $s$, $\sigma_s$ is the standard deviation for all subjects in session $s$ and $\sigma_{si}$ is the standard deviation for subject $i$ in session $s$.  As the files in each session were unique, means and standard deviations were calculated on the subset of files matching those in the session.  Normalized opinion scores were not truncated, however MOS were limited to the subjective interval of 1-5.

% ----------------------------------Subjective Testing Results----------------------------------
\subsection{Results}
A total of 42,529 file ratings were collected from 263 participants across 633 sessions, with 10,354 ratings collected during laboratory testing.  Participants ranged in age from 16 to 66 with a median age of 30.  52.36\% of ratings were contributed by expert listeners.  12 files were limited to a MOS of 1, while 28 files were limited to a MOS of 5.

Due to the different files and time-scale ratios used for the testing subset, direct comparison between methods in training and testing subsets was not appropriate.  However, a general comparison was achieved through local averaging of MOS, centered around training time-scale ratios.  Means of adjacent time-scale ratios, bounded by 0.3 and 3, defined the local areas.  While 0.3 is greater than some time-scales used within the testing set, it was set empirically to include enough data points, while limiting the impact of much slower time-scales.  Mean MOS for testing subset methods are noisier due to the smaller number of files, and non-uniform difficulty in processing each signal.

\begin{figure}[ht]
    \centering
    \includegraphics[trim={0 0 0 0},clip,width=\reprintcolumnwidth]{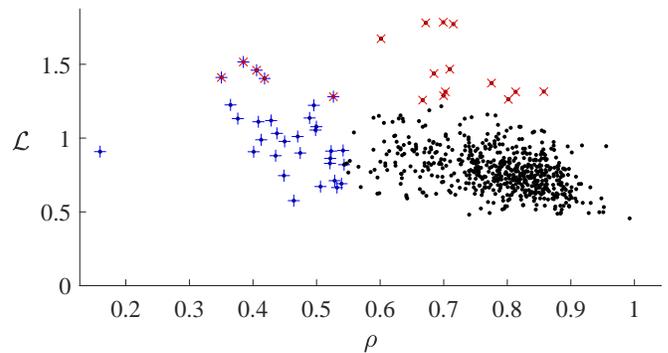}
    \caption{(Color Online) Distribution of PCC and RMSE for all sessions before normalization and outlier removal.  Blue plus symbols mark PCC outliers, while red crosses mark RMSE outliers.}
    \label{fig:Outliers}
\end{figure}

Two measures of reliability were used for each session. The Root Mean Squared Error (RMSE) denoted by $\mathcal{L}$ is given by
\begin{equation}
 \label{eq:RMSE}
 \mathcal{L} =\sqrt{\frac{\sum_{i=1}^{N} \left( \bar{x}_{i}-x_{i} \right) ^2}{N}}
\end{equation}
% \begin{equation}
%  \label{eq:MAD_Mean}
%  \bar{X}_s = \frac{\sum_{i=1}^{N_s}|\bar{x}_{i,s}-x_{i,s}|}{N_{s}}
% \end{equation}
% standard deviation of absolute difference ($\sigma_s$)
% \begin{equation}
%  \label{eq:StdMAD}
%  \sigma_s = \sqrt{\frac{\sum_{i=1}^{N_s}{\big(\bar{X}_s - |\bar{x}_{i,s}-x_{i,s}|\big)}}{N_s-1}}
% \end{equation}
where the number of files within the session is denoted by $N$, $x_{i}$ is the participants opinion score for the file and $\bar{x}_{i}$ is the overall MOS for the file.  The Pearson Correlation Coefficient (PCC), denoted by $\rho$, given by
\begin{equation}
 \label{eq:PCC}
 \rho = \frac{\text{cov}(\bm{x},\bar{\bm{x}})} {\sigma_{\bm{x}} \sigma_{\bar{\bm{x}}}}
\end{equation}
was also used where $\bm{x}$ and $\bar{\bm{x}}$ denote sets of opinion scores and MOS for the session and $\sigma_{\bm{x}}$ and $\sigma_{\bar{\bm{x}}}$ are the standard deviation of $\bm{x}$ and $\bar{\bm{x}}$.  These measures were calculated for each session before and after normalization.   Outliers, calculated prior to normalization and shown in figure \ref{fig:Outliers}, were determined as sessions in which $\mathcal{L}$ or $\rho$ were further than three scaled median absolute deviations away from their respective medians.  This resulted in the removal of 45 sessions containing a total of 2,102 ratings (4.94\%) from the final pool of sessions.

\begin{figure}[ht]
    \centering
    \includegraphics[width=\reprintcolumnwidth]{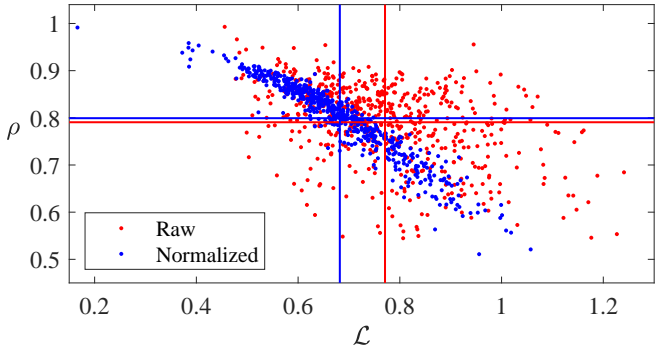}
    \caption{(Color Online) Distribution of PCC and RMSE for each session before normalization.  Horizontal and vertical lines denote means.}
    \label{fig:Pre_Post_Norm}
\end{figure}

Following outlier removal and normalization, $\mathcal{L}$ and $\rho$ means of 0.771 and 0.791 improved to 0.682 and 0.799. Distributions of $\mathcal{L}$ and $\rho$ pre- and post-normalization can be seen in figure \ref{fig:Pre_Post_Norm}.

%\citep{Shrout_Fleiss_1979, Loizou_2013}
The use of Intraclass Correlation Coefficients (ICC) was explored, however as the subjective results are neither fully crossed nor fully nested, ICC cannot be used.  Instead, the interrater reliability for Ill-Structured Measurement Designs of \citet{Putka_Le_2008} was used, calculated by
\begin{equation}
 \label{eq:ISMD_G}
 G(q,k)=\frac{\hat{\sigma}^2_T}{\hat{\sigma}^2_T+ \Big( q\hat{\sigma}^2_R+\frac{\hat{\sigma}^2_{TR,e}}{\hat{k}}\Big)}
\end{equation}
where $\hat{\sigma}^2_T$ is the estimated variance for file main effects (true score), $\hat{\sigma}^2_R$ is the estimated variance for participant main effects, $\hat{\sigma}^2_{TR,e}$ is the estimated variance components for the combination of residual effects and file-participant interaction, and $\hat{k}$ is the harmonic mean of the number of participants per file. $q$ scales the contribution of $\hat{\sigma}^2_R$ based on the overlap between the sets of participants who rate each file, and is calculated by
\begin{equation}
 \label{eq:ISMD_q}
q=\frac{1}{\hat{k}}-\frac{\sum_i{\sum_{i^\prime}{\frac{c_{i,i^\prime}}{k_i k_{i^\prime}}}}}{N_t (N_t-1)}
\end{equation}
where $c_{i,i^\prime}$ is the number of participants that each pair of files ($i, i^\prime$) share, $k_i$ and $k_i{^\prime}$ are the number of participants who rated files $i$ and $i^\prime$ respectively and $N_t$ is the total number of participants in the sample.  This measure gives an overall rater reliability ($G(q,k)$) of 0.871 prior to normalization and 0.909 post normalization.

\begin{figure}[ht]
    \centering
    \includegraphics[width=\reprintcolumnwidth]{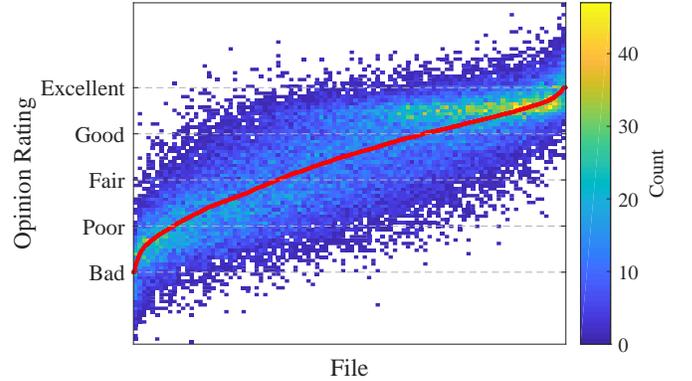}
    \caption{(Color Online) 2D Histogram of normalized responses, ordered by ascending MOS (red line).}
    \label{fig:All_Responses_Mean}
\end{figure}

For an overview of all results, figure \ref{fig:All_Responses_Mean} shows all normalized file ratings ordered by ascending MOS.  All opinion scores are shown in the histogram with the overlaid red line showing the MOS for each file.  It can be seen that when the TSM quality is very high or very low there is greater consensus amongst participants, however there is a large variance in opinion for files with mid-range quality. It can also be seen that the MOS tracks below the majority of responses in the Good to Excellent range, suggesting a difference between MOS and a majority of opinion scores.  Median opinions scores were explored, based on \citep{Jamieson_2004}, resulting in tighter groupings, however there was no significant change in averaged scores nor improvement in session reliability.  Median opinion scores have nonetheless been included as labels with the dataset, along with mean and median opinion scores calculated before normalization.

\begin{figure}[ht]
    \centering
    \includegraphics[width=\reprintcolumnwidth]{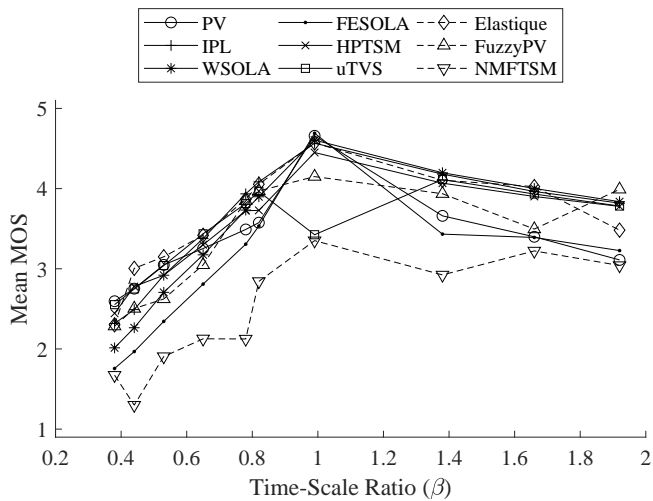}
    \caption{Overall means for each method at each time-scale for all evaluated files.}
    \label{fig:Overall_Means}
\end{figure}

All methods show improvement in quality as $\beta$ approaches 1, as is to be expected.  However, the implementation of uTVS gave poor performance when time-scaling at 0.9961, see section \ref{sec:Dataset}, but achieved state-of-the-art performance for all other time-scales.  Figure \ref{fig:Overall_Means} shows the results of each method for each time-scale, averaged across all files.  When comparing two inverse time-scale ratios, for example $\beta=0.5$ and $\beta=2$, the slower of the pair is lower in quality, suggesting that slowing a file down is perceptually more difficult than increasing its speed.  This is consistent with the testing of \citet{Sharma_2017}, however the effect is more pronounced within this testing.  Of interest are two specific cases, that of PV and WSOLA.  For $\beta<1$, PV is perceived to have a higher quality than WSOLA, however this is reversed for $\beta>1$.  It can then be inferred that different artefacts are perceived as having a greater impact on the quality of the TSM.  We propose that for $\beta<1$, the transient-doubling of WSOLA is perceived as worse than the `phasiness' and transient smearing of the PV, while for $\beta>1$ transient skipping is less detrimental than the artefacts introduced by the PV. This is a similar finding to \citet{Moinet_Dutoit_2011}, who noted that some listeners preferred PV artefacts in some cases.  Similarly, comparison of PV and IPL shows a change in preference towards the smeary PV artefacts for large reductions in speed, over the metallic artefacts of IPL.  The PV was rated comparably to state-of-the-art methods for the three smallest $\beta$.

A surprising result is the high performance of IPL in comparison to HPTSM and uTVS.  HPTSM achieved numerically similar results to those given in \citet{Driedger_Muller_Ewert_2014}.  However, while HPTSM was shown to be greater in MOS by 1, our testing found IPL to be rated higher for all except the two slowest time-scale ratios.  Artefacts due to harmonic-percussive separation, the use of WSOLA with a very short frame length or the lower sample-rate of the files used in the MATLAB TSM Toolbox may be the cause.  Similarly, the reduced sample-rate in original uTVS testing may have contributed to the variance in MOS between testing.  Future research should include comparisons between different IPL implementations.

% \subsection{Per Source File Class}
% , shown in table \ref{tab:MOS_Results},
Algorithm performance per class generally follows that of the overall results.  As expected however, there are differences in performance quality between methods dependent on the source material.  When the mean MOS for each class are considered and $\beta=0.9961$ results excluded, uTVS is preferred for music and solo instrument sources while WSOLA is preferred for voice sources.  However, the differences in averaged ratings are minor in most cases.  Exact mean results have not been reported here as the primary focus is rating time-scaled files, rather than definitive evaluation of different TSM methods.

% 99.61 percent results removed.
% \begin{table}
% \caption{Mean MOS for Overall and Music, Solo Instrument, Voice classes of training source file.  MOS for $\beta=0.9961$ excluded.}
% \label{tab:MOS_Results}
% \centering
% \begin{ruledtabular}
% \begin{tabular}{ccccc}
%  & \textbf{Music} & \textbf{Solo} & \textbf{Voice} & \textbf{Overall} \\
% \hline
% \textbf{PV} & 3.450 & 3.291 & 2.886 & 3.202 \\
% \hline
% \textbf{IPL} & 3.537 & 3.636 & 3.190 & 3.453 \\
% \hline
% \textbf{WSOLA} & 3.133 & 3.547 & \textbf{3.262} & 3.323 \\
% \hline
% \textbf{FESOLA} & 2.418 & 3.203 & 2.968 & 2.882 \\
% \hline
% \textbf{HPTSM} & 3.453 & 3.643 & 3.118 & 3.406 \\
% \hline
% \textbf{uTVS} & \textbf{3.583} & \textbf{3.719} & 3.159 & \textbf{3.486} \\
% \end{tabular}
% \end{ruledtabular}
% \end{table}

%ALL Results
% \begin{table}[ht]
% \centering
% \caption{Mean MOS for each class of training source file, Music, Solo Instrument, Voice, Overall classes are considered.}
% \label{tab:MOS_Results}
% \begin{ruledtabular}
% \begin{tabular}{ccccc}
% %\hline
%  & \textbf{Music} & \textbf{Solo} & \textbf{Voice} & \textbf{Overall} \\
% \hline
% \textbf{PV} & 3.562 & 3.428 & 3.072 & 3.348 \\
% \hline
% \textbf{IPL} & \textbf{3.647} & \textbf{3.730} & 3.320 & \textbf{3.564} \\
% \hline
% \textbf{WSOLA} & 3.273 & 3.658 & \textbf{3.399} & 3.451 \\
% \hline
% \textbf{FESOLA} & 2.636 & 3.355 & 3.146 & 3.063 \\
% \hline
% \textbf{HPTSM} & 3.560 & 3.729 & 3.238 & 3.510 \\
% \hline
% \textbf{uTVS} & 3.575 & 3.684 & 3.184 & 3.480 \\
% %\hline
% \end{tabular}
% \end{ruledtabular}
% \end{table}

\begin{figure}[ht]
    \centering
    \includegraphics[width=\reprintcolumnwidth]{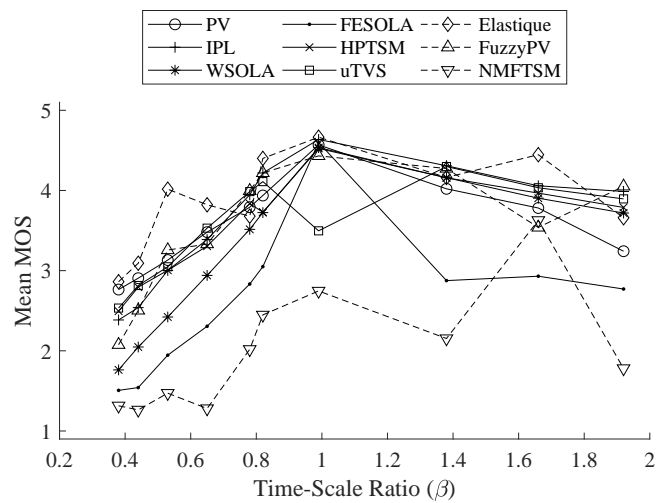}
    \caption{Mean MOS for each method at each time-scale for musical source material.}
    \label{fig:Music_Means}
\end{figure}

Perception of processing quality for musical sources, figure \ref{fig:Music_Means}, confirms the lower quality of time-domain methods, with FESOLA and WSOLA giving poor results.  The most interesting result here is that the PV is consistently rated higher than other methods for $\beta<0.7$ and is comparable for other $\beta$.  If ratings are averaged for each source file, it is possible to identify `difficult' files to process.  Files with uncorrelated high frequency content were rated poorly, while clean, harmonically simple musical excerpts were rated highly.  Signals containing more transient material were rated lower than less transient material. Mean ratings ranged from 2.76 for \textit{Jazz\_1.wav} to 3.94 for \textit{Yellow\_2.wav}.

\begin{figure}[ht]
    \centering
    \includegraphics[width=\reprintcolumnwidth]{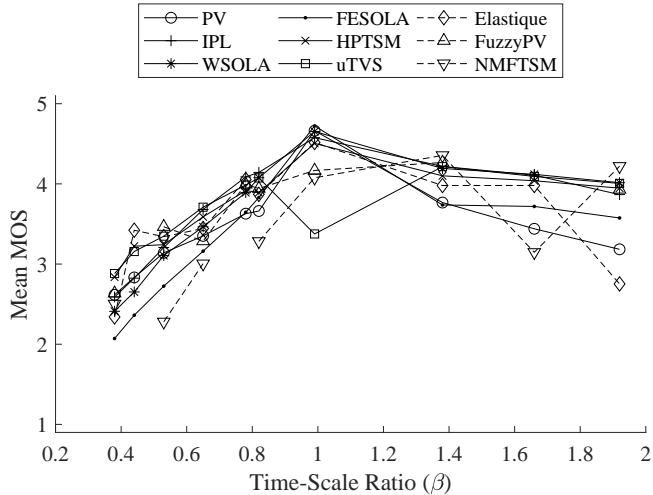}
    \caption{Mean MOS for each method at each time-scale for solo instrument source material.}
    \label{fig:Solo_Means}
\end{figure}

Mean MOS results for the solo instrument class of signals, shown in figure \ref{fig:Solo_Means}, improve over musical and voice classes with the exception of the PV for $\beta>1$.  Synthesizer bass sounds were the lowest rated, followed by noisy percussion, polyphonic instruments and tuned percussion, with monophonic harmonic instruments rated highest.  The combination of low frequencies with significant transients within the synthesizer bass was particularly troublesome for all TSM methods.  Mean file ratings ranged from 2.54 for \textit{Synth\_Bass\_1.wav} to 4.17 for \textit{Ocarina\_01.wav}.

\begin{figure}[ht]
    \centering
    \includegraphics[width=\reprintcolumnwidth]{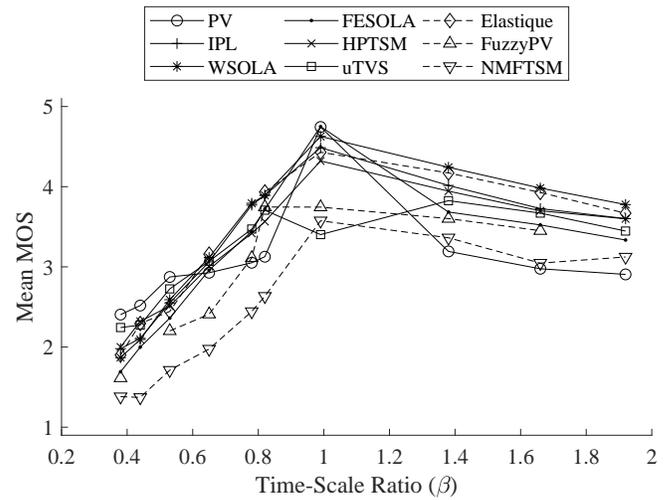}
    \caption{Mean MOS for each method at each time-scale for Voice source material.}
    \label{fig:Voice_Means}
\end{figure}

In considering mean MOS for voice signals, shown in figure \ref{fig:Voice_Means}, WSOLA is preferred for $\beta>1$, while the preference is less clear for $\beta<1$.  Most methods, except the PV and NMFTSM, were rated similarly for $0.6<\beta<1$, however the PV is clearly preferred for $\beta<0.6$.  After this point, smoothness is preferred over transient doubling and metallic artefacts.  When considering mean file ratings, the 11 lowest rated files were all male voices, with female and child voices as the seven highest rated files.  This mirrors results by \citet{Sylvestre_Kabal_1992} who suggested poor frequency resolution for lower frequencies as well as short frame sizes as causes for lower quality.  Mean file ratings ranged from 2.73 for \textit{Male\_18.wav} to 3.59 for \textit{Child\_01.wav}.
% The smoothness of the testing subset methods is due to the homogeneous nature of speech signals, in comparison to the other classes.

\begin{figure}[ht]
    \centering
    \includegraphics[width=\reprintcolumnwidth]{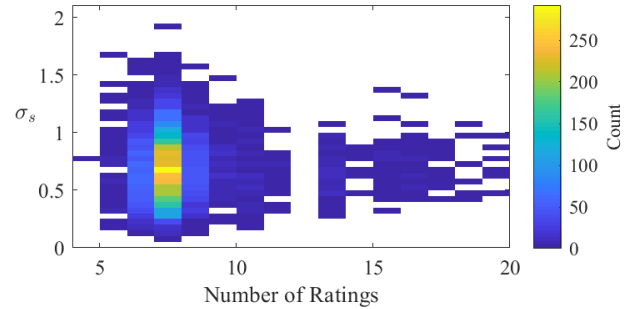}
    \caption{(Color Online) MOS standard deviation against the number of responses for that file.}
    \label{fig:STD_per_responses}
\end{figure}

% \subsection{Analysis of File Standard Deviation}
The mean standard deviation across all files was 0.802 and 0.718, before and after normalization respectively.  As can be seen in figure \ref{fig:STD_per_responses}, the range of standard deviation values converges as the number of responses for the file increases.  During testing (around 19,000 ratings) this graph showed convergence at around seven ratings per file.  As a result, a minimum of seven ratings per file was set as the target to give a `true' representation of the quality of the audio file.  While there are files that have yet to converge, this is a small subset of the total dataset.

% \subsection{Participant Analysis}
Comparisons between expert and non-expert listeners, participants with and without known hearing issues and testing modalities were undertaken using the two one sided tests (TOST) of \citet{Hauck_1984} and \citet{Lakens_2017}.  TOST begins with the null hypothesis of non-equivalent means and uses two one sided tests to show equivalence within a given interval.  The interval can be given as a raw score or a standardized difference.  If the confidence interval for the difference of the means falls within the equivalence interval, the null hypothesis is rejected and equivalence can be claimed.  Analysis was undertaken on session RMSE and PCC values before normalization.  The equivalence interval was calculated at 5\% of the reference sample's mean and Confidence Intervals (CI) of 95\% were used throughout.  Cohen's sample \textit{d} is also given for indication of effect size, where d $\approx$ 0.2 is a small effect size.

\begin{figure}[ht]
    \centering
    \includegraphics[width=\reprintcolumnwidth]{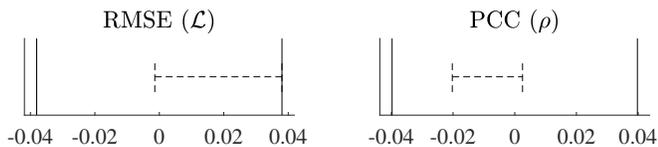}
    \caption{TOST (1-$\alpha$)100\% CI for equivalence of participant experience for $\alpha=0.05$. Equivalence interval of $\pm$5\% of expert participant means.}
    \label{fig:Expert_Non_Expert}
\end{figure}

ITU Recommendation BS.1284 \citep{ITU_BS1284} recommends investigation of the relationship between expert and non-expert listeners.  Participants selected if they had experience critically evaluating the quality of audio.  RMSE and PCC for non-expert listeners were found to be equivalent to those of expert listeners, with equivalence intervals shown in figure \ref{fig:Expert_Non_Expert}.  Testing RMSE gave a maximum p value of 0.0498 and \textit{d} of 0.1273. Testing PCC gave a maximum p value of 4.67e-06 and \textit{d} of 0.1059.  We propose that equivalence is a result of the reference-test style of testing and the medium to large impairment in the processed signal, reducing the importance of highly trained critical listening skills for this type of subjective testing.

\begin{figure}[ht]
    \centering
    \includegraphics[width=\reprintcolumnwidth]{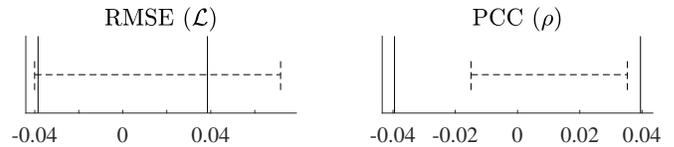}
    \caption{TOST (1-$\alpha$)100\% CI for equivalence of means of participants with and without hearing issues for $\alpha=0.05$. Equivalence interval of $\pm$5\% of mean for participants without hearing issues.}
    \label{fig:Hearing_vs_MAD}
\end{figure}

Participants also reported any known hearing issues, with an open answer text box given for responses.  Results were not excluded if known issues were reported, but were instead manually sorted into a binary classification of `No known hearing issues' and `Any known hearing issues'.  Hearing issues included highly descriptive explanations such as ``-6dB above 14kHz'', a range of tinnitus severity, age related hearing changes and ``I like punk music''.  PCC for participants with any hearing issues were found to be equivalent to those without issue, while RMSE was not found to be equivalent. Equivalence intervals are shown in figure \ref{fig:Hearing_vs_MAD}.  Testing RMSE gave a maximum p value of 0.2467 and \textit{d} of 0.0958. Testing PCC gave a maximum p value of 0.0245 and \textit{d} of 0.1219.  Our proposed explanation is two-fold.  Those participants who reported known hearing issues in great detail were also expert listeners, and familiar with the shortcomings of their own auditory system.  Additionally, as the participants were presented with the source and processed files and asked to rate the quality of the processing, any issue within the auditory system would affect perception of both files.  The small number of sessions classified as `any issue', 33 compared to 554 for `no issue', also impacts this result, greatly increasing the standard error.  A t-test applied to RMSE was unable to reject that the means are equal with a p-score of 0.4985.  Increasing the equivalence interval to $\pm$9.32\% allows RMSE equivalence to be claimed.  Due to the strong PCC equivalence and close RMSE equivalence, we find no reason to reject sessions in which hearing issues were reported.

\begin{figure}[ht]
    \centering
    \includegraphics[width=\reprintcolumnwidth]{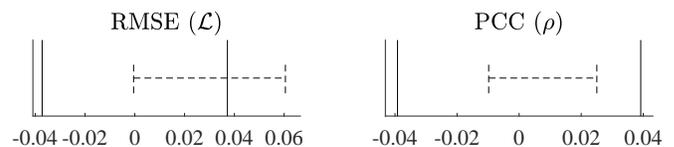}
    \caption{TOST (1-$\alpha$)100\% CI for equivalence of testing modality means for $\alpha=0.05$. Equivalence interval of $\pm$5\% of laboratory participant means.}
    \label{fig:Offline_vs_Online}
\end{figure}

As testing was undertaken in different modalities, comparative analysis of results is necessary.  PCC for remote participants were found to be equivalent to laboratory participants, while RMSE was not found to be equivalent. Equivalence intervals are shown in figure  \ref{fig:Offline_vs_Online}.  Testing RMSE gave a maximum p value of 0.3474 and \textit{d} of 0.2126. Testing PCC gave a maximum p value of 0.0013 and \textit{d} of 0.0931.  A t-test applied to RMSE was unable to reject that the means are equal with a p-score of 0.4693.  Increasing the equivalence interval to $\pm$8.14\% allowed RMSE equivalence to be claimed.  Due to the strong PCC equivalence and close RMSE equivalence, we found no reason to reject either testing mode.

Analysis of the possible impact of age on the quality of the participant's responses was undertaken.  Correlations of 0.108 and -0.001 were found between the age of the participant and the RMSE or PCC respectively, showing no impact of age on evaluation ability.

% Pearson Correlation of age to RMSE = -0.00103211
% Pearson Correlation of age to PCC = 0.108587

% \begin{figure}[ht]
%     \centering
%     \includegraphics[width=\reprintcolumnwidth]{./Age_vs_RMSE_norm.eps}
%     \caption{Comparison of RMSE ($\mathcal{L}$) and participant age.}
%     \label{fig:Age_vs_MAD}
% \end{figure}

% \subsection{Recommendations}
% \label{sec:Recommendations}
% Based on this research a number of recommendations can be presented to assist future subjective testing.  The number of files presented in each session should be kept as small as practically possible, particularly for inexperienced listeners.  We found 60 to 80 files to be optimal in terms of participation and completion of sessions.  The ease of use of the testing software should also be considered.  Recruiting participants became easier once using the online solution. Additionally, using the level of impairment qualifier could be a better choice than Bad to Excellent due to the large artefacts present in TSM.

% \subsection{Availability}
% \label{sec:Availability}
The labeled dataset is available, under the Creative Commons Attribution 4.0 International (CC BY 4.0) license, through IEEE-Dataport at http://ieee-dataport.org/1987. Implementation and additional source code is available at github.com/zygurt/TSM.

\section{Towards an Objective Measure of Quality}
\label{sec:Objective}
% \subsection{Comparison with Objective Measures}
Comparison between MOS and previous objective measures, $SER$ and $D_M$, found correlations of 0.3707 and 0.1574 respectively by averaging absolute correlations for $\beta<1$ and $\beta>1$.  Signals were aligned through time axis interpolation of the reference magnitude spectrum to the duration of the test spectrum.

% \subsection{Objective evaluation using a modified PEAQ implementation}
Perceptual Evaluation of Audio Quality (PEAQ) \citep{Thiede_2000,ITU_BS1387_PEAQ} is often used for objective quality evaluation.  PEAQ extracts perceptually informed features, using differences between reference and test signals, that are fed into a small neural network to predict subjective scores. Direct application to time-scaled signals is not possible however, due a loss of alignment during TSM.  Initial testing, applying the dataset in the design of an objective measure of quality, was undertaken using a modified version of PEAQ.  Signals were aligned as above and gave similar correlation to MOS as $SER$ and $D_M$.  The original PEAQ basic neural network was retrained to the subjective MOS, with 10\% of the training set reserved for validation. Training used seeds of 0 to 99, with the optimal epoch given by the minimum overall distance ($\mathcal{D}$)
\begin{equation}
\label{eq:Overall_Distance1}
    \mathcal{D} = \|[\hat{\rho}, \hat{\mathcal{L}}]\|_{_2}
\end{equation}
where $\hat{\rho}$ and $\hat{\mathcal{L}}$ are calculated by
\begin{equation}
\label{eq:Overall_Distance2}
    \hat{\rho} = \|[1-\overline{\bm{\rho}},(max(\bm{\rho})-min(\bm{\rho}))]\|_{_2}
\end{equation}
\begin{equation}
\label{eq:Overall_Distance3}
    \hat{\mathcal{L}} = \|[\overline{\bm{\mathcal{L}}},(max(\bm{\mathcal{L}})-min(\bm{\mathcal{L}}))]\|_{_2}
\end{equation}
where $\bm{\rho}=[\rho_{tr}, \rho_{val}, \rho_{te}]$, $\bm{\mathcal{L}}=[\mathcal{L}_{tr}, \mathcal{L}_{val}, \mathcal{L}_{te}]$ and $tr$, $val$ and $te$ denote training, validation and testing.  The best network achieved a $\mathcal{D}$ of 0.731 and an $\overline{\bm{\mathcal{L}}}$ of 0.668 and $\overline{\bm{\rho}}$ of 0.719, placing it at the 11th and 17th percentiles of subjective sessions.

\begin{figure}[ht]
    \centering
    %trim= {40 0 35 10},clip,
    \includegraphics[width=\reprintcolumnwidth]{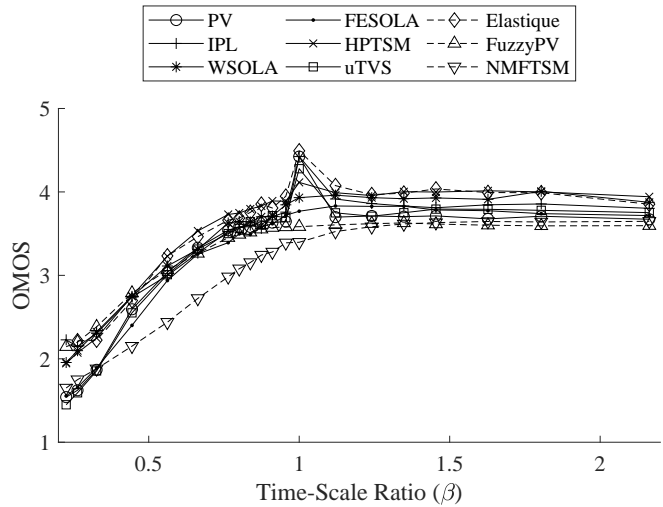}
    \caption{Objective MOS for each method in the evaluation set, averaged at each time-scale ratio.}
    \label{fig:PEAQ_OMOQ}
\end{figure}

An evaluation set was created by processing the testing subset source files with all methods previously mentioned, at 20 time-scale ratios in the range of $0.22<\beta<2.2$.  The mean objective output for each method across the range of time-scales is shown in figure \ref{fig:PEAQ_OMOQ}.

The output exhibits a similar shape to the subjective results, however it only moves away from the mean for $\beta<0.75$ and $\beta=1$.  Development of an accurate objective measure of quality for TSM algorithms is now achievable, and the aim of future work.
% ----------------------------------Conclusions----------------------------------
\section{Conclusion}
\label{sec:Conclusion}
This paper detailed the creation, subjective evaluation and analysis of a dataset and its use in the development of an objective measure of quality for time-scaled audio.  Six TSM methods processed 88 source files at 10 time-scales resulting in 5,280 processed signals for a training subset. Three additional methods at four random time-scales resulted in 240 signals for a testing subset.  42,529 ratings were collected from 633 sessions using laboratory and remote collection methods.  Preliminary results for an objective measure of quality were presented, which achieved an RMSE loss of 0.668 and PCC of 0.719.  The aim of future work is the design of an improved objective measure of quality for TSM using the dataset, to assist in comparative evaluation of novel methods.

\begin{acknowledgments}
The authors would like to acknowledge and thank all of the participants who assisted in the subjective evaluation of the dataset as well as the reviewers for their comments that significantly improved the paper.
\end{acknowledgments}
\bibliography{SubjTSMDB}

\begin{thebibliography}{29}
\def\enquote#1{``#1,''}
\def\plainquote#1{``#1''}
\expandafter\ifx\csname natexlab\endcsname\relax\def\natexlab#1{#1}\fi
\providecommand{\dourl}[1]{\href{http://#1}{\nolinkurl{#1}}}
\providecommand{\bibinfo}[2]{#2}
\providecommand{\noopsort}[1]{}
\providecommand{\switchargs}[2]{#2#1}
  \def\eatspace #1{#1}

\bibitem[{Damsk{\"a}gg and V{\"a}lim{\"a}ki(2017)}]{Damskagg_2017}
\bibinfo{author}{Damsk{\"a}gg, E.},  and \bibinfo{author}{V{\"a}lim{\"a}ki, V.}
  (\textbf{\bibinfo{year}{2017}}). \enquote{\bibinfo{title}{Audio time
  stretching using fuzzy classification of spectral bins}}
  \bibinfo{journal}{Applied Sciences} \textbf{7}(12), \bibinfo{pages}{1293}.

\bibitem[{Driedger and Muller(2014)}]{Driedger_Muller_2014}
\bibinfo{author}{Driedger, J.},  and \bibinfo{author}{Muller, M.}
  (\textbf{\bibinfo{year}{2014}}). \enquote{\bibinfo{title}{{TSM} toolbox:
  {MATLAB} implementations of time-scale modification algorithms}} in
  \emph{\bibinfo{booktitle}{Proc. of the 17th Int. Conference on Digital Audio
  Effects (DAFx-14)}}, \bibinfo{address}{Erlangen, Germany}, pp.
  \bibinfo{pages}{1--8}.

\bibitem[{Driedger \emph{et~al.}(2014)Driedger, Muller, and
  Ewert}]{Driedger_Muller_Ewert_2014}
\bibinfo{author}{Driedger, J.}, \bibinfo{author}{Muller, M.},  and
  \bibinfo{author}{Ewert, S.} (\textbf{\bibinfo{year}{2014}}).
  \enquote{\bibinfo{title}{Improving time-scale modification of music signals
  using harmonic-percussive separation}} \bibinfo{journal}{IEEE Signal
  Processing Letters} \textbf{21}(1), \bibinfo{pages}{105--109}.

\bibitem[{Hauck and Anderson(1984)}]{Hauck_1984}
\bibinfo{author}{Hauck, W.~W.},  and \bibinfo{author}{Anderson, S.}
  (\textbf{\bibinfo{year}{1984}}). \enquote{\bibinfo{title}{A new statistical
  procedure for testing equivalence in two-group comparative bioavailability
  trials}} \bibinfo{journal}{Journal of Pharmacokinetics and Biopharmaceutics}
  \textbf{12}(1), \bibinfo{pages}{83--91}.

\bibitem[{ITU-T(2001)}]{ITU_BS1387_PEAQ}
\bibinfo{author}{ITU-T} (\textbf{\bibinfo{year}{2001}}).
  \enquote{\bibinfo{title}{Itu-r bs. 1387-1: Method for objective measurements
  of perceived audio quality}} \bibinfo{type}{Technical Report}.

\bibitem[{ITU-T(2019)}]{ITU_BS1284}
\bibinfo{author}{ITU-T} (\textbf{\bibinfo{year}{2019}}).
  \enquote{\bibinfo{title}{Itu-r bs. 1284-1: General methods for the subjective
  assessment of sound quality}} \bibinfo{type}{Technical Report}.

\bibitem[{Jamieson \emph{et~al.}(2004)}]{Jamieson_2004}
\bibinfo{author}{Jamieson, S.} \emph{et~al.} (\textbf{\bibinfo{year}{2004}}).
  \enquote{\bibinfo{title}{Likert scales: how to (ab) use them}}
  \bibinfo{journal}{Medical education} \textbf{38}(12),
  \bibinfo{pages}{1217--1218}.

\bibitem[{Jillings \emph{et~al.}(2015)Jillings, Moffat, De~Man, and
  Reiss}]{WAET_2015}
\bibinfo{author}{Jillings, N.}, \bibinfo{author}{Moffat, D.},
  \bibinfo{author}{De~Man, B.},  and \bibinfo{author}{Reiss, J.~D.}
  (\textbf{\bibinfo{year}{2015}}). \enquote{\bibinfo{title}{Web {A}udio
  {E}valuation {T}ool: {A} browser-based listening test environment}} in
  \emph{\bibinfo{booktitle}{12th Sound and Music Computing Conference}},
  \bibinfo{address}{Maynooth, Ireland}, pp. \bibinfo{pages}{147--152}.

\bibitem[{Karrer \emph{et~al.}(2006)Karrer, Lee, and
  Borchers}]{Karrer_Lee_Borchers_2006}
\bibinfo{author}{Karrer, T.}, \bibinfo{author}{Lee, E.},  and
  \bibinfo{author}{Borchers, J.} (\textbf{\bibinfo{year}{2006}}).
  \enquote{\bibinfo{title}{{PhaVoRIT}: A phase vocoder for real-time
  interactive time-stretching}} \bibinfo{type}{Technical Report}.

\bibitem[{Lakens(2017)}]{Lakens_2017}
\bibinfo{author}{Lakens, D.} (\textbf{\bibinfo{year}{2017}}).
  \enquote{\bibinfo{title}{Equivalence tests: A practical primer for t tests,
  correlations, and meta-analyses}} \bibinfo{journal}{Social psychological and
  personality science} \textbf{8}(4), \bibinfo{pages}{355--362}.

\bibitem[{Laroche(2002)}]{Laroche_2002}
\bibinfo{author}{Laroche, J.} (\textbf{\bibinfo{year}{2002}}).
  \emph{\bibinfo{title}{Time and Pitch Scale Modification of Audio Signals}},
  \bibinfo{pages}{279--309} (\bibinfo{publisher}{Springer US},
  \bibinfo{address}{Boston, MA}), \dodoi{10.1007/0-306-47042-X_7}.

\bibitem[{Laroche and Dolson(1997)}]{Laroche_Dolson_1997}
\bibinfo{author}{Laroche, J.},  and \bibinfo{author}{Dolson, M.}
  (\textbf{\bibinfo{year}{1997}}). \enquote{\bibinfo{title}{Phase-vocoder:
  About this phasiness business}} in \emph{\bibinfo{booktitle}{Proceedings of
  1997 Workshop on Applications of Signal Processing to Audio and Acoustics}},
  \bibinfo{organization}{IEEE}, pp. \bibinfo{pages}{1--4}.

\bibitem[{Laroche and Dolson(1999)}]{Laroche_Dolson_1999}
\bibinfo{author}{Laroche, J.},  and \bibinfo{author}{Dolson, M.}
  (\textbf{\bibinfo{year}{1999}}). \enquote{\bibinfo{title}{Improved phase
  vocoder time-scale modification of audio}} \bibinfo{journal}{IEEE
  Transactions on Speech and Audio Processing} \textbf{7}(3),
  \bibinfo{pages}{323--332}.

\bibitem[{Moinet and Dutoit(2011)}]{Moinet_Dutoit_2011}
\bibinfo{author}{Moinet, A.},  and \bibinfo{author}{Dutoit, T.}
  (\textbf{\bibinfo{year}{2011}}). \enquote{\bibinfo{title}{{PVSOLA}: A phase
  vocoder with synchronized overlap-add}} in \emph{\bibinfo{booktitle}{Proc. of
  the 14th Int. Conference on Digital Audio Effects (DAFx-11)}},
  \bibinfo{address}{Paris, France}, pp. \bibinfo{pages}{269--275}.

\bibitem[{Moulines and Charpentier(1990)}]{Moulines_1990}
\bibinfo{author}{Moulines, E.},  and \bibinfo{author}{Charpentier, F.}
  (\textbf{\bibinfo{year}{1990}}). \enquote{\bibinfo{title}{Pitch-synchronous
  waveform processing techniques for text-to-speech synthesis using diphones}}
  \bibinfo{journal}{Speech communication} \textbf{9}(5-6),
  \bibinfo{pages}{453--467}.

\bibitem[{Moulines and Laroche(1995)}]{Moulines_Laroche_1995}
\bibinfo{author}{Moulines, E.},  and \bibinfo{author}{Laroche, J.}
  (\textbf{\bibinfo{year}{1995}}). \enquote{\bibinfo{title}{Non-parametric
  techniques for pitch-scale and time-scale modification of speech}}
  \bibinfo{journal}{Speech communication} \textbf{16}(2),
  \bibinfo{pages}{175--205}.

\bibitem[{Portnoff(1976)}]{Portnoff_1976}
\bibinfo{author}{Portnoff, M.} (\textbf{\bibinfo{year}{1976}}).
  \enquote{\bibinfo{title}{Implementation of the digital phase vocoder using
  the fast {F}ourier transform}} \bibinfo{journal}{IEEE Transactions on
  Acoustics, Speech, And Signal Processing} \textbf{24}(3),
  \bibinfo{pages}{243--248}.

\bibitem[{Portnoff(1981)}]{Portnoff_1981}
\bibinfo{author}{Portnoff, M.} (\textbf{\bibinfo{year}{1981}}).
  \enquote{\bibinfo{title}{Time-scale modification of speech based on
  short-time {F}ourier analysis}} \bibinfo{journal}{IEEE Transactions on
  Acoustics, Speech, and Signal Processing} \textbf{29}(3),
  \bibinfo{pages}{374--390}.

\bibitem[{Putka \emph{et~al.}(2008)Putka, Le, McCloy, and Diaz}]{Putka_Le_2008}
\bibinfo{author}{Putka, D.~J.}, \bibinfo{author}{Le, H.},
  \bibinfo{author}{McCloy, R.~A.},  and \bibinfo{author}{Diaz, T.}
  (\textbf{\bibinfo{year}{2008}}). \enquote{\bibinfo{title}{Ill-structured
  measurement designs in organizational research: Implications for estimating
  interrater reliability}} \bibinfo{journal}{Journal of Applied Psychology}
  \textbf{93}(5), \bibinfo{pages}{959}.

\bibitem[{Roberts and Paliwal(2018)}]{Roberts_2018_Stereo}
\bibinfo{author}{Roberts, T.},  and \bibinfo{author}{Paliwal, K.~K.}
  (\textbf{\bibinfo{year}{2018}}). \enquote{\bibinfo{title}{Stereo time-scale
  modification using sum and difference transformation}} in
  \emph{\bibinfo{booktitle}{2018 12th International Conference on Signal
  Processing and Communication Systems (ICSPCS)}},
  \bibinfo{organization}{IEEE}, pp. \bibinfo{pages}{1--5}.

\bibitem[{Roberts and Paliwal(2019)}]{Roberts_2019_FESOLA}
\bibinfo{author}{Roberts, T.},  and \bibinfo{author}{Paliwal, K.~K.}
  (\textbf{\bibinfo{year}{2019}}). \enquote{\bibinfo{title}{Time-scale
  modification using fuzzy epoch-synchronous overlap-add ({FESOLA})}} in
  \emph{\bibinfo{booktitle}{2019 IEEE Workshop on Applications of Signal
  Processing to Audio and Acoustics}}, \bibinfo{organization}{IEEE}, pp.
  \bibinfo{pages}{31--34}.

\bibitem[{Roma \emph{et~al.}(2019)Roma, Green, and
  Tremblay}]{Roma_Green_Tremblay_2019}
\bibinfo{author}{Roma, G.}, \bibinfo{author}{Green, O.},  and
  \bibinfo{author}{Tremblay, P.} (\textbf{\bibinfo{year}{2019}}).
  \enquote{\bibinfo{title}{Time scale modification of audio using non-negative
  matrix factorization}} in \emph{\bibinfo{booktitle}{Proc. of the 22nd Int.
  Conference on Digital Audio Effects (DAFx-19)}},
  \bibinfo{address}{Birmingham, UK}, pp. \bibinfo{pages}{1--6}.

\bibitem[{Roucos and Wilgus(1985)}]{Roucos_Wilgus_1985}
\bibinfo{author}{Roucos, S.},  and \bibinfo{author}{Wilgus, A.}
  (\textbf{\bibinfo{year}{1985}}). \enquote{\bibinfo{title}{High quality
  time-scale modification for speech}} in \emph{\bibinfo{booktitle}{Proceedings
  of ICASSP '85}}, \bibinfo{organization}{IEEE}, Vol. 10, pp.
  \bibinfo{pages}{493--496}.

\bibitem[{Sharma \emph{et~al.}(2017)Sharma, Potadar, Chetupalli, and
  Sreenivas}]{Sharma_2017}
\bibinfo{author}{Sharma, N.}, \bibinfo{author}{Potadar, S.},
  \bibinfo{author}{Chetupalli, S.~R.},  and \bibinfo{author}{Sreenivas, T.}
  (\textbf{\bibinfo{year}{2017}}). \enquote{\bibinfo{title}{Mel-scale sub-band
  modelling for perceptually improved time-scale modification of speech and
  audio signals}} in \emph{\bibinfo{booktitle}{2017 Twenty-third National
  Conference on Communications (NCC)}}, \bibinfo{organization}{IEEE}, pp.
  \bibinfo{pages}{1--5}.

\bibitem[{Sylvestre and Kabal(1992)}]{Sylvestre_Kabal_1992}
\bibinfo{author}{Sylvestre, B.},  and \bibinfo{author}{Kabal, P.}
  (\textbf{\bibinfo{year}{1992}}). \enquote{\bibinfo{title}{Time-scale
  modification of speech using an incremental time-frequency approach with
  waveform structure compensation}} in \emph{\bibinfo{booktitle}{Proceedings of
  ICASSP '92}}, \bibinfo{organization}{IEEE}, Vol. 1, pp.
  \bibinfo{pages}{81--84}.

\bibitem[{Thiede \emph{et~al.}(2000)Thiede, Treurniet, Bitto, Schmidmer,
  Sporer, Beerends, and Colomes}]{Thiede_2000}
\bibinfo{author}{Thiede, T.}, \bibinfo{author}{Treurniet, W.~C.},
  \bibinfo{author}{Bitto, R.}, \bibinfo{author}{Schmidmer, C.},
  \bibinfo{author}{Sporer, T.}, \bibinfo{author}{Beerends, J.~G.},  and
  \bibinfo{author}{Colomes, C.} (\textbf{\bibinfo{year}{2000}}).
  \enquote{\bibinfo{title}{Peaq-the itu standard for objective measurement of
  perceived audio quality}} \bibinfo{journal}{Journal of the Audio Engineering
  Society} \textbf{48}(1/2), \bibinfo{pages}{3--29}.

\bibitem[{Torcoli(2019)}]{Matteo_2019}
\bibinfo{author}{Torcoli, M.} (\textbf{\bibinfo{year}{2019}}).
  \enquote{\bibinfo{title}{An improved measure of musical noise based on
  spectral kurtosis}} in \emph{\bibinfo{booktitle}{2019 IEEE Workshop on
  Applications of Signal Processing to Audio and Acoustics}},
  \bibinfo{organization}{IEEE}, pp. \bibinfo{pages}{90--94}.

\bibitem[{Verhelst and Roelands(1993)}]{Verhelst_Roelands_1993}
\bibinfo{author}{Verhelst, W.},  and \bibinfo{author}{Roelands, M.}
  (\textbf{\bibinfo{year}{1993}}). \enquote{\bibinfo{title}{An overlap-add
  technique based on waveform similarity ({WSOLA}) for high quality time-scale
  modification of speech}} in \emph{\bibinfo{booktitle}{Proceedings of ICASSP
  '93}}, \bibinfo{organization}{IEEE}, Vol. 2, pp. \bibinfo{pages}{554--557}.

\bibitem[{{Zplane Development}(2018)}]{elastique}
\bibinfo{author}{{Zplane Development}} (\textbf{\bibinfo{year}{2018}}).
  \plainquote{\bibinfo{title}{\`{E}lastique time stretching \& pitch shifting
  sdks (version 3.2.5) [computer program]}}
  \dourl{http://licensing.zplane.de/technology\#elastique},
  \bibinfo{note}{(Last viewed October 31, 2019)}.

\end{thebibliography}
\end{document}